\documentclass[12pt]{article}
\usepackage[numbers,compress]{natbib}
\begin{document}

\input amssym.tex

\title{Rest frame vacuum of the Proca field on the de Sitter expanding universe}

\author{Ion I. Cot\u aescu  \\
{\small \it West University of Timi\c soara,}\\
{\small \it V. Parvan Ave. 4 RO-300223 Timi\c soara,  Romania}}

\maketitle

\begin{abstract}
The general plane wave solutions of the Proca field in conformal charts of the de Sitter expanding universe are derived for arbitrary polarizations showing how the frequencies can be separated in rest frames, defining thus the rest frame vacuum of this field. 

 Pacs: 04.62.+v
\end{abstract}



\section{Introduction}

The quantum modes of the Proca field \cite{Proca}  (i. e. the massive charged vector field) on the de Sitter  space-time are less studied since the method of the maximaly symmetric two-point functions  \cite{PPV,PPV1} opened an optimistic alternative horizon some time ago. Thus we know so far only the spherical modes derived in spherically symmetric static charts \cite{V} and our plane wave solutions of the momentum-helicity basis \cite{CProca} derived in the comoving chart of conformal time of the de Sitter expanding portion.  

The principal problem of the plane waves on the de Sitter manifold is the frequencies separation for any massive field, regardless its spin.  This comes from the fact that the plane wave mode functions are eigenfunctions of the momentum operator which does not commute with the Hamiltonian (or energy) operator which might determine the frequency as the sign of its eigenvalue. Under such circumstances the criterion of frequencies separation  must be introduced by a supplemental physical assumption.   One says that in this manner one defines the vacuum. The principal method applied so far is to look for the asymptotic mode functions whose behavior is similar to the usual Minkowskian particle and antiparticle mode functions as in the case of the adiabatic vacua of the Bunch-Davies type  \cite{BuD}.  

However, recently we observed that in the rest frames, where the momentum vanishes, the energy operator commutes with the momentum one such that the frequencies can be separated in these frames just as in special relativity. With this procedure we defined a new type of vacuum called the rest frame vacuum we applied to the Dirac \cite{CDrfv} and Klein-Gordon \cite{CKGrfv} fields. In this paper we would like to complete this collection with the rest frame vacuum of the Proca field determining the mode functions which become energy eigenstates in the rest frames.  

For this purpose we need to know the most general form of the vector plane waves with any polarization but this is not studied till now since only the helicity basis was used in applications. For this reason we derive first the general form of the vector plane waves of any polarization and then we write down  the mode functions of the rest frame vacuum.  Both these results are reported here for the first time.

The paper is organized as follows. We start in the second section presenting the principal features of the Proca theory on the de Sitter expanding universe.  The next section is devoted to the general plane wave solution of any polarization deriving suitable sets of orthonormal modes able to constitute different generalised bases. In the third section we show how the frequencies can be separated in rest frames, defining thus the rest frame vacuum for any polarization. Finally we present our concluding remarks.

\section{Proca field on de Sitter expanding universe}

Let us start with the expanding portion of the $(1+3)$-dimensional de Sitter space-time denoting by $\omega $ the Hubble de Sitter constant since in our notation $H$ is reserved for the Hamiltonian operator.  We use the standard notations with natural indices, $\alpha, \beta,...\mu,\nu...=0,1,2,3$ as well as the vector notation for the space vectors.  Here we consider only the conformal chart $\{x^{\mu}\}=\{t_c,{\bf x}\}$ with the {conformal} time $t_c$ and Cartesian space coordinates ${\bf x}=(x^1,x^2,x^3)$ having the line element \cite{BD}
\begin{equation}
ds^{2}=g_{\mu \nu}d{x}^{\mu}d{x}^{\nu}=\frac{1}{(\omega t)^2}\,\eta_{\mu \nu} dx^{\mu}
dx^{\nu}=\frac{1}{(\omega t)^2}\,\left(dt_c^{2}- d{\bf x}\cdot d{\bf
x}\right)\,,
\end{equation}
where $\eta={\rm diag}(1,-1,-1,-1)$ is the the Minkowski metric which will be useful in further calculations. On the de Sitter expanding portion, the conformal time is negative, $t_c\in (-\infty,0]$,   being related to the proper (or cosmic) time $t\in(-\infty,\infty)$ as $t=-\frac{1}{\omega}\ln(-\omega t_c)$. 

The de Sitter manifold is a spatially flat space-time of maximal symmetry \cite{SW} having the $SO(1,4)$ isometry group  whose generators  are associated to ten independent Killing vectors fields. These give rise to the basis-generators of the vector representation of the $SO(1,4)$ group carried by the space of the vector fields $A$ \cite{CProca}. In what follows  we need to use only on the Hamiltonian (or energy) operator $H$  the momentum components $P^i$,
\begin{eqnarray}
(H\, A)_{\mu}&=&-i\omega(t_c \partial_{t_c} + x^i\partial_i +1
) {A}_{\mu}\,,\label{HH}\\
(P^i A)_{\mu}&=&-i\partial_i\,A_{\mu}\,.\label{PP}
\end{eqnarray}
that satisfy the commutation rule  \cite{CProca,CGRG},
\begin{equation}\label{HPP}
\left[ H, P^i\right]=i\omega P_i \,,
\end{equation}
but commute with the operator of the field equation.

In the chart $\{t_c,{\bf x}\}$ the equation of the  Proca free field of mass $m$, minimally coupled  to the de Sitter gravity, read \cite{CProca},
\begin{eqnarray}
&&\partial_{t_c}(\partial_i A_i)-\Delta
A_0+\frac{\mu^2}{t_c^2}\,A_0=0 \,,\label{EdS1}\\
&&\partial_{t_c}^2A_k-\Delta A_k-\partial_k(\partial_{t_c} A_0)+\partial_k(\partial_i
A_i)+\frac{\mu^2}{t_c^2}\,A_k=0\,,\label{EdS2}
\end{eqnarray}
where $\mu=\frac{m}{\omega}$. Hereby it results that the Lorentz condition,  
\begin{equation}\label{Lor}
\partial_i A_i =\partial_{t_c} A_0-\frac{2}{t_c}\,A_0\,,
\end{equation}
is mandatory assuring the uniqueness of the spin $s=1$.
The solutions of these equations must be normalized (in usual or generalized sense) with respect to the relativistic scalar product that in the conformal chart takes the simpler form \cite{CProca}, 
\begin{equation}\label{AA}
\left<A|\,A'\right>=-i\eta^{\mu\nu}\int
d^3x\, A^*_{\mu}(t,{\bf x})
\stackrel{\leftrightarrow\,\,\,}{\partial_{t_c}} A'_{\nu}(t,{\bf x}) \,.
\end{equation}
laying out  the following Hermitian properties,
\begin{equation}\label{AAs}
\langle A|\,A'\rangle=\langle A'|\,A\rangle^*=-\langle A^*|\,{A'}^*\rangle=-\langle {A'}^*|\,{A}^*\rangle^*\,,
\end{equation}
resulted from the definition (\ref{AA}).

The 'squared norms' $\langle A| A\rangle$ of the square integrable solutions $A\in {\cal H}\subset{\cal A}$ can take any real value  splitting  the whole space of solutions,  ${\cal A}$, as
\begin{equation}
A \in \left\{ 
\begin{array}{lll}
{\cal H}_+\subset{\cal A}_+& {\rm if}& \langle A|\,A\rangle>0\,,\\
{\cal H}_0\subset{\cal A}_0& {\rm if}& \langle A|\,A\rangle=0\,,\\
{\cal H}_-\subset{\cal A}_-& {\rm if}& \langle A|\,A\rangle<0\,.\\
\end{array}
\right.
\end{equation} 
From the physical point of view  the mode functions of  ${\cal A}_{\pm}$ are of positive/negative frequencies while those of  ${\cal A}_0$ do not  have a physical meaning. Given  $A\in {\cal A}_+$ then  $A^*\in {\cal A}_-$ satisfies  $\langle A^*|\, A^*\rangle =-\langle A|\, A\rangle$ and $\langle A^*|\, A\rangle=0$ which means that  $A$ and $A^*$ are orthogonal each other. Consequently, a linear combination of normalized solutions, $\hat A =c_1A+c_2 A^*$, may have any 'squared norm' since $\langle \hat A|\,\hat A\rangle=|c_1|^2-|c_2|^2$. In particular, whether  $A^*=A$ then  $A\in {\cal A}_0$ since then $\langle A|\, A\rangle=0$ as it results from Eq. ({\ref{AAs}).   Concluding we can say that in fact,  ${\cal H}$ is a Krein space while ${\cal A}_{\pm}$ are the spaces of  tempered distributions of the Hilbertian triads associated to the Hilbert spaces ${\cal H}_{\pm}$ equipped with the scalar products $\pm \langle~|~\rangle$. 

For deriving concrete solutions of the Proca equations we need a complete system of commuting operators which might determine the solutions as common eigenfunctions whose eighenvalues should play the role of  integration constants.  Unfortunately, the Cartan algebra of the $SO(1,4)$ group is too poor for offering us complete systems of commuting operators \cite{CGRG}. In the case of the plane wave solutions we have to use the incomplete system $\{P^1,P^2,P^3\}$ since the Hamiltonian operator commutes with this system only in  rest frames as it results  from Eq. ({\ref{HPP}).    

\section{General plane wave solutions}

The solutions of the Proca equations can be expanded as,
\begin{eqnarray}
A_{\mu}(t_c,{\bf x})&=&A^{(+)}_{\mu}(t_c,{\bf x})+A^{(-)}_{\mu}(t_c,{\bf x})\nonumber\\
&=&\int d^3p \sum_{\lambda}\left[A_{({\bf p},{\bf e}_{\lambda})\,\mu}(t_c,{\bf x}) a({\bf p},\lambda)+
A_{({\bf p},{\bf e}_{\lambda})\,\mu}^*(t_c,{\bf x}) b^*({\bf p},\lambda)\right]\,,\label{field1}
\end{eqnarray}
in terms of wave functions in momentum representation, $a({\bf p},\lambda)$,
and $b({\bf p},\lambda)$,  and the fundamental solutions of positive frequency $A_{({\bf p},{\bf e}_{\lambda})\,\mu}(t_c,{\bf x})$ assumed to be eigenfunctions of the momentum operators, $P^i A_{({\bf p},{\bf e}_{\lambda)}\mu}=p^i A_{({\bf p},{\bf e}_{\lambda})\mu}$. In addition, these depend on  an arbitrary polarization given by the polarization unit vectors ${\bf e}_{\lambda}$ which can be specified at any time according to our needs. These solutions must satisfy the generalized  orthonormalization relations
\begin{equation}\label{orto}
\langle A_{({\bf p},{\bf e}_{\lambda})}|\, A_{({\bf p}',{\bf e}_{\lambda'})}\rangle=\delta_{\lambda\lambda'}\delta^3({\bf p}-{\bf p}')\,,
\end{equation}
of the generalized orthonormal basis $\{A_{({\bf p},{\bf e}_{\lambda})} \}\subset{\cal A}_+$.

For investigating the structure of these functions  we consider an arbitrary real valued  unit vector ${\bf e}$ separating the space part as  
\begin{equation}\label{Acc}
A_{({\bf p},{\bf e})\mu}(t_c,{\bf x})={\rm f}_{({\bf p},{\bf e})\mu}(t_c) \frac{e^{i({\bf p}\cdot{\bf x})}}{(2\pi)^{\frac{3}{2}}}\,,
\end{equation}
where ${\rm f}_{({\bf p},{\bf e})\mu}(t_c)$  are time modulation functions. Then the scalar product        
\begin{equation}
\langle A_{({\bf p},{\bf e})}|\, A_{({\bf p}',{\bf e})}\rangle=\delta^3({\bf p}-{\bf p}')\rangle\left[-i\eta^{\mu\nu}{\rm f}_{({\bf p},{\bf e})\mu}^*(t_c)\stackrel{\leftrightarrow\,\,\,}{\partial_{t_c}} {\rm f}_{({\bf p},{\bf e})\nu}(t_c)\right]\,,\label{Aff}
\end{equation}
complies with  the orthonormalization condition (\ref{orto})  only if the time modulation functions satisfy 
\begin{equation}\label{ff}
({\rm f}_{({\bf p},{\bf e})},{\rm f}_{({\bf p},{\bf e})})=-i\eta^{\mu\nu}{\rm f}_{({\bf p},{\bf e})\mu}^*(t_c)\stackrel{\leftrightarrow\,\,\,}{\partial_{t_c}} {\rm f}_{({\bf p},{\bf e})\nu}(t_c)=1\,.
\end{equation} 
Furthermore,  we separate the transverse part, orthogonal to the momentum direction, as 
\begin{eqnarray}
{\rm f}_{({\bf p},{\bf e}) i}(t_c)&=&\left(e^i-\frac{p^i}{p}\, ({\bf e}\cdot{\bf n}_p)\right){\cal F}_p(t_c)+\frac{p^i}{p}\,({\bf e}\cdot{\bf n}_p)\,{\cal K}_p(t_c) \,, ~~ i=1,2,3\,,~~~~\label{fff1}\\
{\rm f}_{({\bf p},{\bf e})0}(t_c)&=&({\bf e}\cdot{\bf n}_p)\,{\cal H}_p(t_c) \label{fff2}
\end{eqnarray}
denoting $p=|{\bf p}|$, ${\bf n}_p=\frac{{\bf p}}{p}$ while ${\cal F}_p$, ${\cal K}_p$ and ${\cal H}_p$ are new time-dependent functions. The function ${\cal F}_p$ gives the time modulation of the transverse part  while ${\cal K}_p$ and ${\cal H}_p$ govern the longitudinal part, along  ${\bf n}_p$. 

Looking for analytical solutions we substitute the form (\ref{Acc}) in Eqs. (\ref{EdS1}) and (\ref{EdS2}) finding that only the functions  ${\cal F}_p$ and ${\cal H}_p$ are independent, satisfying the second order equations  
\begin{eqnarray}
&&\frac{d^2{\cal F}_p(t_c)}{dt_c^2}+\left(\frac{\mu^2}{t_c^2}+p^2\right) {\cal F}_p(t_c)=0\label{EqF}\,,\\
&&\frac{d^2{\cal H}_p(t_c)}{dt^2}-\frac{2}{ t_c }\frac{d{\cal H}_p(t_c)}{dt_c}+\left(\frac{\mu^2+2}{ t_c^2}+p^2\right) {\cal H}_p(t_c)=0\,. \label{EqK}
\end{eqnarray}
The third equation we need  is given  just by  the Lorentz condition (\ref{Lor})  as
\begin{equation}\label{EqH}
{\cal K}_p(t_c)=-\frac{i}{p}\left(\frac{d{\cal H}_p(t_c)}{dt_c}-\frac{2}{t_c}{\cal H}_p(t_c) \right) \,.
\end{equation}
Eqs. (\ref{EqF}) and (\ref{EqK}) can be solved in terms of modified Bessel functions $K$ obtaining the solutions  
\begin{eqnarray}
{\cal F}_p(t_c)&=&\alpha N f_p(t_c)\,, \quad f_p(t_c)= \sqrt{-t_c}\, K_{\nu}(ipt_c)\,,\label{Ff}\\
{\cal H}_p(t_c)&=&\beta N h_p(t_c)\,, \quad h_p(t_c)= \frac{2 ip}{1-2\nu}(-t_c)^{\frac{3}{2}}K_{\nu}(ipt_c)\,,\label{Hh}
\end{eqnarray}
depending on the free parameters $\alpha, \beta \in {\Bbb C}$ while $N$ is the general normalization factor. The indices of the modified Bessel functions take the values 
\begin{equation}\label{ndS}
\nu=\left\{\begin{array}{lll}
\sqrt{\frac{1}{4}-\mu^2}&{\rm for} & \mu<\frac{1}{2}\\
i\kappa\,,\quad \kappa= \sqrt{\mu^2-\frac{1}{4}}&{\rm  for} & \mu>\frac{1}{2}
\end{array} \right. \,.
\end{equation}
In addition, from Eq. (\ref{EqH}) we obtain 
\begin{equation}
{\cal K}_p(t_c)=\beta N k_p(t_c)\,,
\end{equation}
where
\begin{equation}\label{Kk}
 k_p(t_c)= \sqrt{-t_c}\, K_{\nu}(ipt_c) -\frac{2ip}{1-2\nu}(-t)^{\frac{3}{2}}K_{1+\nu}(ipt_c) \,.
\end{equation}
In what follows we say that these are $K$-solution for distinguish them from other solutions expressed  in terms of different Bessel functions. 

It remains to impose the normalization condition (\ref{ff}) considering separately the cases of $m<\frac{1}{2}\omega$ and $m>\frac{1}{2}\omega$. Fortunately, by using Eq. (\ref{KuKu}) we find an analytic normalization formula that holds  on both these domains giving the normalization factor  
\begin{equation}\label{NNor}
N_{({\bf p},{\bf e})}^{(\alpha,\beta)}=\frac{1}{\sqrt{\pi}}\left[|\alpha|^2\left(1-({\bf e}\cdot{\bf n}_p)^2\right)+|\beta|^2({\bf e}\cdot{\bf n}_p)^2\left|\frac{1+2\nu}{1-2\nu}\right|\right]^{-\frac{1}{2}}\,,
\end{equation}
which, in general, may depend on  polarization and momentum direction. We observe that in the domain $\mu>\frac{1}{2}$, where $\nu=i\kappa$ as in Eq. (\ref{ndS}), the quantity
\begin{equation}\label{eta}
\epsilon=\frac{1-2i\kappa}{1+2 i\kappa}\,, 
\end{equation}
is a phase factor, with $ |\epsilon|=1$.

Now we have all the elements for writing down the general form of the fundamental
solutions of positive frequency 
\begin{equation}\label{Acc}
A_{({\bf p},{\bf e})\mu}^{(\alpha,\beta)}(t_c,{\bf x})= {\rm f}_{({\bf p},{\bf e})\mu}^{(\alpha,\beta)}(t_c) \frac{e^{i({\bf p}\cdot{\bf x})}}{(2\pi)^{\frac{3}{2}}}
\end{equation}
depending on the time  modulation functions 
\begin{eqnarray}
{\rm f}_{({\bf p},{\bf e})i}^{(\alpha,\beta)}(t_c)&=& N_{({\bf p},{\bf e})}^{(\alpha,\beta)} \left\{ \left[\alpha \,e^i +(\beta-\alpha)\frac{p^i}{p}({\bf e}\cdot{\bf n}_p)\right]\sqrt{-t_c}\,K_{\nu}(ipt_c))\right.\nonumber\\
&&\hspace*{18mm}\left. -\,\beta\, p^i \frac{2 i }{1-2\nu}({\bf e}\cdot{\bf n}_p)(-t_c)^{\frac{3}{2}}K_{1+\nu}(ipt_c) \right\}\,,\label{Sfin1}\\
{\rm f}_{({\bf p},{\bf e})0}^{(\alpha,\beta)}(t_c)&=& N_{({\bf p},{\bf e})}^{(\alpha,\beta)} \,\beta\, \frac{2 i p}{1-2\nu}({\bf e}\cdot{\bf n}_p)(-t_c)^{\frac{3}{2}}K_{\nu}(ipt_c) \,.\label{Sfin2}
\end{eqnarray}
In particular, the transverse solution $A_{({\bf p},{\bf e})}^{(1,0)}$ and the longitudinal one $A_{({\bf p},{\bf e})}^{(0,1)}$ form an orthonormal system satisfying
\begin{eqnarray}
&&\left<A_{({\bf p},{\bf e})}^{(1,0)}| A_{({\bf p}',{\bf e})}^{(1,0)}\right>=
\left<A_{({\bf p},{\bf e})}^{(0,1)}| A_{({\bf p}',{\bf e})}^{(0,1)}\right>=\delta^3({\bf p}-{\bf p}')\,,\\
&&\left<A_{({\bf p},{\bf e})}^{(1,0)}| A_{({\bf p}',{\bf e})}^{(0,1)}\right>=
\left<A_{({\bf p},{\bf e})}^{(1,0)}| A_{({\bf p}',{\bf e})}^{(0,1)\,*}\right>=0\,,
\end{eqnarray}
since the quantities, 
\begin{equation}
\left( {\rm f}_{({\bf p},{\bf e})}^{(1,0)}\,,\, {\rm f}_{({\bf p}',{\bf e})}^{(0,1)}\right) =
\left( {\rm f}_{({\bf p},{\bf e})}^{(1,0)}\,,\, {\rm f}_{({\bf p}',{\bf e})}^{(0,1)\,*}\right)=0\,.
\end{equation} 
vanish as orthogonal four vectors with respect to the Minkowski metric $\eta$. Another interesting case is when $\alpha=\beta$ since then the second term of Eq. (\ref{Sfin1}) vanishes and for $\mu >\frac{1}{2}$ we have
\begin{equation}
N_{({\bf p},{\bf e})}^{(\alpha,\alpha)}=\frac{1}{\sqrt{\pi} |\alpha|}~~\to~~A_{({\bf p},{\bf e})}^{(\alpha,\alpha )}=\frac{\alpha}{|\alpha|}A_{({\bf p},{\bf e})}^{(1,1 )}\,.
\end{equation}

All the above general results are obtained for a polarization unit vector ${\bf e}$ fixed in an arbitrary direction.  In practice one uses the helicity basis  $\{{\bf e}_{\lambda}({\bf n}_p) | \lambda= 0,  \pm 1\}$ (presented in the Appendix B) whose unit vectors are related to the momentum direction such that 
\begin{equation}
{\bf e}_{\lambda}({\bf n}_p)\cdot{\bf n}_p = \left\{
\begin{array}{lll}
0& {\rm for}& \lambda=\pm 1\\
1& {\rm for}& \lambda=0
\end{array}\right. \,.
\end{equation}
Consequently, the solutions $A_{({\bf p},\lambda)}^{(\alpha,\beta)}\equiv A_{({\bf p},{\bf e}_{\lambda})}^{(\alpha,\beta)}$ differ only through phase factors from the standard solutions $A_{({\bf p}, \pm1)}\equiv A_{({\bf p}, \pm1)}^{(1,0)}$ and $A_{({\bf p}, 0)}\equiv A_{({\bf p}, 0)}^{(0,1)}$ which satisfy 
\begin{equation}
\left< A_{({\bf p},\lambda)}|A_{({\bf p}',\lambda')}\right>=\delta_{\lambda,\lambda'}\delta^3({\bf p}-{\bf p}')\,,
\end{equation}
forming the momentum-helicity basis $\{ A_{({\bf p},\lambda)} | {\bf p}\in {\Bbb R}^3_p, \lambda=0,\pm1\}\subset {\cal A}_+$. Indeed, according to our general formulas (\ref{NNor}), (\ref{Sfin1}) and (\ref{Sfin2}) we find that 
\begin{equation}
A_{({\bf p}, \pm1)}^{(\alpha,\beta)}=\frac{\alpha}{|\alpha|}A_{({\bf p}, \pm1)}\,,\quad
A_{({\bf p}, 0)}^{(\alpha,\beta)}=\frac{\beta}{|\beta|}A_{({\bf p}, 0)}\,.
\end{equation}
For example, in Ref. \cite{CProca} we derived a particular solution of this type \footnote{In Ref. \cite{CProca} a misprint  must be corrected reading $\left(ik-\frac{1}{2}\right)$ instead of $\left(ik+\frac{1}{2}\right)$ in Eq. (36). This does not affect other results since this equation is not used explicitly.} for $\mu>\frac{1}{2}$, expressed in terms of Hankel functions, with
\begin{equation}
\alpha=1\,, \quad \beta=\frac{1-2i\kappa}{2\mu}~ \to~ N=\frac{1}{\sqrt{\pi}}\,,
\end{equation}
since $|\beta|=1$. This solution corresponds to an adiabatic vacuum, analogous to the Bunch-Davies one of the scalar field \cite{BuD}. Another type of vacuum will be studied  in the next section.

\section{Rest frame vacuum}

We focus now on the rest frame vacuum that can be defined by separating the frequencies in the rest frames, as we proceeded in the case of the Dirac \cite{CDrfv} and Klein-Gordon \cite{CKGrfv} fields. We start with the observation that the energy operator (\ref{HH}) which, in general, does not commute with the momentum operators, takes in the rest frame the form \cite{CProca}
\begin{equation}
H=-i\omega(t_c \partial_{t_c} + 1)=i\partial_t-i\omega\,,
\end{equation}
since here $\partial_i A=p^i A=0$. In other respects, we know that on the de Sitter spacetime, for $m>\frac{1}{2}\omega $, the positive ($+$) or negative ($-$) rest energies are \cite{CGRG}
\begin{equation}
E^{\pm}_0=\pm\omega\kappa-\frac{3i\omega}{2}=\pm\hat m -\frac{3i\omega}{2}\,,
\end{equation}
where $\hat m=\omega\kappa$ is the rest energy interpreted as an effective (or dynamical) mass while the imaginary term is due to the de Sitter expansion \cite{CGRG}. This means that for separating the frequencies in the rest frame we must look for time modulation functions $\hat {\rm f}_p^{\pm}$ which satisfy the eigenvalue problems $H \hat {\rm f}_0^{\pm}=E_0^{\pm}\hat {\rm f}_0^{\pm}$ in this frame taking the form, 
\begin{equation}\label{f0tc}
\hat {\rm f}_0^{\pm}(t_c)=\lim_{p\to 0}{\rm f}_p^{\pm}(t_c) \propto \sqrt{t_c}\,{t_c}^{\pm i\kappa}\,. 
\end{equation}
The above  eigenvalues problems and the obvious property $\hat {\rm f}_0^-=(\hat {\rm f}_0^+)^*$ indicate that the functions  of positive frequencies,  associated to particle states, must behave in rest frames as $\hat {\rm f}_0^+$  while those of negative  frequencies, describing antiparticle states, as $\hat {\rm f}_0^-$. With this guide we can separate the frequencies in the rest frames selecting  time modulation functions with suitable properties.

Bearing in mind that only the modified Bessel functions $I$ may have the convenient behavior (\ref{I0}) in the rest frame for $\mu>\frac{1}{2}$, we consider the new $I$-solutions of positive frequency of the form (\ref{fff1}) and (\ref{fff2}) whose functions, 
\begin{equation}
\hat {\cal F}_p(t_c)=\hat\alpha \hat N\hat f_p(t_c)\,,\quad \hat {\cal H}_p(t_c)=\hat\beta \hat N\hat h_p(t_c)\,, \quad \hat {\cal K}_p(t_c)=\hat\beta\hat  N \hat k_p(t_c)
\end{equation}
depend on the new parameters  $\hat\alpha$ and $\hat\beta$  and functions
\begin{eqnarray}
\hat f_p(t_c)&=& \sqrt{-t_c}\, I_{i\kappa}(ipt_c)\,,\label{fhk1}\\
\hat h_p(t_c)&=& \frac{2 ip}{1-2i\kappa}(-t_c)^{\frac{3}{2}}I_{i\kappa}(ipt_c)\,,\label{fhk2}\\
\hat k_p(t_c)&=& \sqrt{-t_c}\, I_{i\kappa}(ipt_c) +\frac{2ip}{1-2i\kappa}(-t_c)^{\frac{3}{2}}I_{1+i\kappa}(ipt_c)\,.\label{fhk3}
\end{eqnarray}
We obtain thus time modulation functions having  similar structures as in Eqs. (\ref{Sfin1}) and (\ref{Sfin2}), laying out similar terms, but depending on the new functions (\ref{fhk1}),(\ref{fhk2}) and (\ref{fhk3}). We must stress that the limit for ${\bf  p}\to 0$ of the term proportional with  $(\hat\alpha-\hat\beta)\frac{p^i}{p}I_{i\kappa}(ipt_c)$  remains undetermined. Therefore,  we must drop it out by taking $\hat\alpha=\hat\beta=1$,  restricting ourselves only to solutions of the form  $\hat A_{({\bf p},{\bf e})}^{(1,1)}$ which will be denoted from now simply as $\hat A_{({\bf p},{\bf e})}$.

The new  normalization constant $\hat N$ is different from that calculated in the previous section since the functions $I$ satisfy the identity (\ref{IuIu}) which is different from that of the functions $K$,  (\ref{KuKu}).  Then, after a little calculation, we may write
\begin{equation}
 \hat N=\sqrt{\frac{\pi}{e^{2\pi\kappa}-1}}\left(\frac{p}{2\omega}\right)^{-i\kappa}\,,
\end{equation}
fixing the general phase factor we need for assuring the correct limit of the functions $I_{i\kappa}(ipt_c)$ which  behaves as in Eq. (\ref{I0}) when  $p\to 0$. Thus we arrive at the final form of the fundamental solutions of positive frequency in the rest frame vacuum,
\begin{eqnarray}
\hat A^>_{({\bf p},{\bf e}) i}(t_c,{\bf x})&=& \hat N\, \frac{e^{i({\bf p}\cdot{\bf x})}}{(2\pi)^{\frac{3}{2}}}\left[ e^i \sqrt{-t_c}\,I_{i\kappa}(ipt_c)\right.\nonumber\\
&&\hspace*{18mm} \left.+\,{p^i} \frac{2 i ({\bf e}\cdot{\bf n}_p)}{1-2i\kappa}(-t_c)^{\frac{3}{2}}I_{1+i\kappa}(ipt_c) \right]\,, \label{Aifin} \\
\hat A^>_{({\bf p},{\bf e}) 0}(t_c,{\bf x})&=&\hat N\, \frac{e^{i({\bf p}\cdot{\bf x})}}{(2\pi)^{\frac{3}{2}}}\,\frac{2 i ({\bf e}\cdot{\bf p})}{1-2i\kappa}(-t_c)^{\frac{3}{2}}I_{i\kappa}(ipt_c)  \,,~~~~\label{A0fin}
\end{eqnarray} 
that holds only for  $\mu>\frac{1}{2}$. In the limit of ${\bf p}\to 0$ the time-like component (\ref{A0fin}) and the second term of Eq. (\ref{Aifin}) vanish while the limit of the remaining term is, up to a phase factor, of the form
\begin{equation}
\lim_{{\bf p}\to 0}\hat A^>_{({\bf p},{\bf e}) i}(t_c,{\bf x}) \propto \frac{e^i}{(2\pi)^{\frac{3}{2}}}\frac{1}{\sqrt{2 \kappa\omega}}\sqrt{-\omega t_c}(-\omega t_c)^{i\kappa}\,,
\end{equation}
similar to that of the eigenfunction  (\ref{f0tc}). 

Hereby we understand the role of the polarization vector ${\bf e}$ which gives the position of $A$ in the rest frame. By choosing an arbitrary  vector basis or  the spin one (as given in the Appendix B) we can describe the polarization in the rest frame instead of helicity. For example,  we may define the momentum-spin basis of ${\cal A}_+$ formed by the solutions $A^>_{({\bf p},\sigma) \mu}(t_c,{\bf x})\equiv A^>_{({\bf p},{\bf e}_{\sigma}) \mu}(t_c,{\bf x})$. Obviously, this basis can be replaced at any time by the usual helicity one but only now after separating the frequencies in the rest frame since the helicity is related to a non-vanishing momentum. 

In the domain $\mu<\frac{1}{2}$ we obtain similar solutions, $\hat A^<_{({\bf p},{\bf e}) \mu}$,  substituting the index $i\kappa$  by $\nu>0$ in Eqs. (\ref{Aifin}) and (\ref{A0fin}). Then for $p\to 0$ the functions $I_{\nu}(ipt_c)\sim (ipt_c)^{\nu}$ have a tachyonic behavior but  vanishing as ${\cal O}(p^{\nu})$  such that the modulation functions collapse before getting a physical interpretation in the rest frame.  However, this is not surprising since these solutions are of null 'squared norm', 
\begin{equation}
\langle \hat A^<_{({\bf p},{\bf e})}|\, \hat A^<_{({\bf p}',{\bf e})}\rangle=0~~ \to~~ \hat  A^<_{({\bf p},{\bf e})}\in{\cal A}_0\,,
\end{equation}
having no physical meaning.
In other words, only for masses larger than $\frac{1}{2}\,\omega$ there are particle and antiparticle solutions correctly defined including in the rest frames of the de Sitter space-time while under this limit the particles cannot survive. 

Finally, let us see how this $I$-solution can be represented in terms of  $K$-solutions  $A_{({\bf p},{\bf e})}^{(\alpha,\beta)}$  in the domain $\mu>\frac{1}{2}$ where our $I$- solution is well-defined. For this purpose  we use Eq. (\ref{KI}) and the identities invilving the functions defined by Eqs. (\ref{Ff}), (\ref{Hh}) and  (\ref{fhk3}) that read
\begin{eqnarray}
f_p(t_c)^*&=&\sqrt{-t_c}K_{i\kappa}(-ipt_c)\,,\\
-\epsilon^*h_p(t_c)^*&=& \frac{2ip}{1-2i\kappa}(-t_c)^{\frac{3}{2}}K_{i\kappa}(-ipt_c)\,,\\
\epsilon^*k_p(t_c)^*-f_p(t_c)^*&=&\frac{2ip}{1-2i\kappa}(-t)^{\frac{3}{2}} K_{1+i\kappa}(-ipt_c)\,, 
\end{eqnarray}
where $\epsilon$  given by Eq. (\ref{eta}).    Then after a little calculation we obtain the following expansion in terms of orthonormal $K$-solutions,
\begin{equation}
\hat  A^>_{({\bf p},{\bf e})}=c_1 A_{({\bf p},{\bf e})} ^{(1,1)}+c_2 \left[A_{({\bf p},{\bf e})}^{(1,-\epsilon)}\right]^*
\end{equation}
where
\begin{equation}
c_1=i \left(\frac{p}{2\omega}\right)^{i\kappa}\frac{e^{\pi\kappa}}{\sqrt{e^{2\pi\kappa}-1}}\,, \quad c_2=i  \left(\frac{p}{2\omega}\right)^{i\kappa}\frac{1}{\sqrt{e^{2\pi\kappa}-1}}\,.
\end{equation}
satisfy the normalization condition $|c_1|^2-|c_2|^2=1$. 

Notice that the $K$-solutions $A_{({\bf p},{\bf e})}^{(\alpha,\beta)}$ cannot be expressed exclusively as linear combinations of the standard $I$-solutions  $A^>_{({\bf p},{\bf e})}$ and   $(A^>_{({\bf p},{\bf e})})^*$ since the structure of the functions $K$ generates supplemental terms. This indicates that the $K$-solutions cannot be interpreted in the rest frames, remaining only with our standard $I$-solutions (\ref{Aifin}) and (\ref{A0fin}) which have a precised physical meaning.

\section{Concluding remarks}

This paper is mainly a technical piece of work reporting two new results, namely the general form of the plane wave solutions of the Proca field on de Sitter expanding universe and the solutions corresponding to the rest frame vacuum. These solutions are presented in the most general context in which the polarization remains arbitrary. Thus we create a general framework for introducing new bases as, for example, the momentum-spin one in which the polarization can be measured with respect to an arbitrary direction in rest frame.  The relation between the momentum-helicity basis and the momentum-spin one could be of interest in further investigations offering new technical advantages. We prepare thus the Proca field for taking over its role in the quantum field theory on the de Sitter space-times where the Klein-Gordon, Maxwell and Dirac fields are already considered for calculating first order processes in the presence of the de Sitter gravity \cite{Lot1,Lot2,Lot3,R1,R2,R3,A1,A2,CQED,Cr1,Cr2}

From this perspective, it is important to have well-defined particles inclusive in rest frames. Here we have seen that in the minimal coupling  the frequencies separation in rest frames can be done only for the particles with $m>\frac{\omega}{2}$ which have a non-vanishing rest energy interpreted as an effective mass, $\hat m>0$. This behavior is not singular since for the Klein-Gordon field we found a similar phenomenon due to a rest energy which is different from the formal mass of the Klein-Gordon equation  \cite{CKGrfv}.  It is remarkable that  in both these cases the unwanted  tachyonic behaviors (for $m<\frac{1}{2}\,\omega$  in our case and for $m<\frac{3}{2}\,\omega$ for the Klein-Gordon field)  are eliminated in a natural manner as long as the corresponding mode functions have null norms \cite{CKGrfv}. This indicates that the rest frame vacuum is a good starting point for the quantum theory of the Proca field.

On the other hand, it is worth pointing out that, in contrast to  the boson fields, the Dirac field minimally coupled to the de Sitter gravity behaves as  in special relativity, its rest energy being just the mass of the Dirac equation \cite{CGRG}. Moreover, we have shown that this property holds on any spatially flat FLRW space-time  \cite{CDrfv}. This discrepancy between the behavior of  bosons and that of fermions in rest frames  cannot be understood  in the actual stage of the theory but  this may open new directions in developing the quantum theory on curved backgrounds.

\appendix
\setcounter{section}{0}\renewcommand{\thesection}{\Alph{section}}
\setcounter{equation}{0} \renewcommand{\theequation}
{A.\arabic{equation}}

\section{Modified Bessel functions}

The modified Bessel functions $I_{\nu}(z)$ and $K_{\nu}(z)$ are related as \cite{NIST}
\begin{eqnarray}
K_{\nu}(z)&=&K_{-\nu}(z)=\frac{\pi}{2}\frac{I_{-\nu}(z)-I_{\nu}(z)}{\sin\pi \nu}\,,\label{IK}\\
I_{\pm\nu}(z)&=&e^{\mp i\pi\nu}I_{\pm\nu}(-z)\nonumber\\
&=&\frac{i}{\pi}\left[K_{\nu}(-z)-e^{\mp i\pi\nu}K_{\nu}(z)\right]\,.\label{KI}
\end{eqnarray}
Their Wronskians  give  the identities we need for normalizing the mode functions. For $\nu=i\kappa$ we obtain
\begin{equation}
i I_{i\kappa}(i s) \stackrel{\leftrightarrow}{\partial_{s}}I_{-i\kappa}(is)= \frac{2\, {\rm sinh}\,\pi\kappa}{\pi s}\,,\label{IuIu}
\end{equation}
while the identity
\begin{equation}
i K_{\nu}(-i s) \stackrel{\leftrightarrow}{\partial_{s}}K_{\nu}(is)
=\frac{\pi}{|s|}\,,\label{KuKu}
\end{equation}
holds for any $\nu$.

For $|z|\to \infty$ and any $\nu$ we have,
\begin{equation}\label{Km0}
 I_{\nu}(z) \to \sqrt{\frac{\pi}{2z}}e^{z}\,, \quad K_{\nu}(z) \to K_{\frac{1}{2}}(z)=\sqrt{\frac{\pi}{2z}}e^{-z}\,.
\end{equation} 
In the limit of $|z|\to 0$ the functions $I_{\nu}$ behave as  
\begin{equation}\label{I0}
I_{\nu}(z)\sim \frac{1}{\Gamma(\nu+1)} \left(\frac{z}{2}\right)^{\nu}\,,
\end{equation}
while for the functions $K_{\nu}$ we have to use Eq. (\ref{IK}).

\section{Polarization}
\setcounter{equation}{0} \renewcommand{\theequation}
{B.\arabic{equation}}

For describing the polarization in the rest frame we may use an  orthogonal basis of unit vectors ,$\{{\bf e}_1,{\bf e}_2, {\bf e}_3\}$, or the associated canonical spin basis of the unit vectors   $\{{\bf e}_{\sigma}\,|\,\sigma=\pm1,0\}$,  defined as 
${\bf e}_+=\frac{1}{\sqrt{2}}({\bf e}_1 +i {\bf e}_2), {\bf e}_-=\frac{1}{\sqrt{2}}(-{\bf e}_1 +i {\bf e}_2), {\bf e}_0={\bf e}_3$.  
This basis is attached to the rest frame, the polarization $\sigma$ giving the projection of the spin on its third axis. 

In contrast,  in the helicity basis $\{{\bf e}_{\lambda}({\bf n}_p) | \lambda= 0,  \pm 1\}$ the axis of the spin projections is along the momentum direction ${\bf n}_p$. Thus   
for $\lambda=0$ we have ${\bf e}({\bf n}_p,0)={\bf n}_p$, while the unit vectors with $\lambda=\pm 1$ are transverse,  ${\bf p}\cdot{\bf e}({\bf n}_p,\pm 1)=0$. In general,
they have  c-number components which satisfy \cite{SW,BS}
\begin{eqnarray}
{\bf e}({\bf n}_p,\lambda)^*\cdot{\bf e}({\bf
n}_p,\lambda')&=&\delta_{\lambda\lambda'}\,,\\
{\bf e}({\bf n}_p,\lambda)^*\land {\bf e}({\bf
n}_p,\lambda)&=&i \lambda\, {\bf n}_p\,,\label{eee}\\
 \sum_{\lambda}e^i({\bf n}_p,\lambda)^*\,e_j({\bf
n}_p,\lambda)&=&\delta_{ij}\,.\label{tran}
\end{eqnarray}
Note that in Eq. (\ref{NNor}) only the real valued unit vector ${\bf e}({\bf n}_p,0)$ gives a non-vanishing contribution.

\end{document}